\documentclass{jfm}

\usepackage{graphicx}
\usepackage{float}
\usepackage{natbib}
\usepackage{color}
\usepackage{amsmath,amssymb}
\graphicspath{{./}}

\ifCUPmtlplainloaded \else
  \checkfont{eurm10}
  \iffontfound
    \IfFileExists{upmath.sty}
      {\typeout{^^JFound AMS Euler Roman fonts on the system,
                   using the 'upmath' package.^^J}%
       \usepackage{upmath}}
      {\typeout{^^JFound AMS Euler Roman fonts on the system, but you
                   dont seem to have the}%
       \typeout{'upmath' package installed. JFM.cls can take advantage
                 of these fonts,^^Jif you use 'upmath' package.^^J}%
      }
  \else    %
  \fi
\fi

\ifCUPmtlplainloaded \else
  \checkfont{msam10}
  \iffontfound
    \IfFileExists{amssymb.sty}
      {\typeout{^^JFound AMS Symbol fonts on the system, using the
                'amssymb' package.^^J}%
       \usepackage{amssymb}%

      }{}
  \fi
\fi

\ifCUPmtlplainloaded \else
  \IfFileExists{amsbsy.sty}
    {\typeout{^^JFound the 'amsbsy' package on the system, using it.^^J}%
     \usepackage{amsbsy}}
    {\providecommand\boldsymbol[1]{\mbox{\boldmath $##1$}}}
\fi

\newcommand\Rey{\mbox{\textit{Re}}}  

\newsavebox{\astrutbox}
\sbox{\astrutbox}{\rule[-5pt]{0pt}{20pt}}

\newcommand{\DBfinal}[1]{{\color{black}#1}}
\newcommand{\DBnew}[1]{{\color{black}#1}}
\newcommand{\DBmaybe}[1]{{\color{black}#1}}
\newcommand{\BSrevision}[1]{{\color{black}#1}}

\title[Speed and structure of turbulent fronts in pipe flow]{Speed and structure of turbulent fronts in pipe flow}

\author[Baofang Song, Dwight Barkley, Bj\"orn Hof, and Marc Avila]%
{Baofang Song$^{1,2}$\thanks{Email address for correspondence: baofang.song@zarm.uni-bremen.de}\ns,
Dwight Barkley$^3$,
Bj\"orn Hof$^{4}$,
\& Marc Avila$^{1,2}$}

\affiliation{
 $^1$Center of Applied Space Technology and Microgravity, University of Bremen, 28359 Bremen,
  Germany\\
[\affilskip] $^2$Institute of Fluid Mechanics, Friedrich-Alexander-Universit\"at
Erlangen-N\"urnberg, 91058 Erlangen, Germany\\
[\affilskip] $^3$Mathematics Institute, University of Warwick, Coventry CV4
7AL, United Kingdom\\
[\affilskip] $^4$Institute of Science and Technology Austria, 3400 Klosterneuburg,
  Austria}

\pubyear{2013}
\volume{650}
\pagerange{119--126}
\date{?; revised ?; accepted ?. - To be entered by editorial office}
\begin{document}

\maketitle

\begin{abstract} 
Using extensive direct numerical simulations, the dynamics of
laminar-turbulent fronts in pipe flow is investigated for Reynolds numbers \BSrevision{between
$\Rey$=2000 and 5500}.
We here investigate the physical distinction between the fronts of weak and
strong slugs both by analysing the turbulent kinetic energy budget and by
comparing the downstream front motion to the advection speed of bulk turbulent
structures.
Our study shows that weak downstream fronts
travel slower than turbulent structures in the bulk and correspond to decaying
turbulence at the front.  At $\Rey\simeq 2900$ the downstream front speed
becomes faster than the advection speed, marking the onset of strong
fronts. In contrast to weak fronts, turbulent eddies are generated at strong
fronts by feeding on the downstream laminar flow.
Our study also suggests that temporal fluctuations of production and
dissipation at the \BSrevision{downstream} laminar-turbulent front drive the dynamical switches
between the two types of front observed up to $\Rey\simeq 3200$.
\end{abstract}


\section{Introduction}
\label{sec:intro}

In pipe flow, turbulence first appears at relatively low Reynolds numbers in
localised patches known as puffs.  Only at higher Reynolds numbers does
turbulence begin to expand in streamwise extent and eventually render the flow
fully turbulent. Such expanding turbulent regions are known as slugs.  The
structure of puffs and slugs, as well as the transformation process between
them as Reynolds number increases, has been the subject of many experimental
and numerical studies \citep{Lindgren1969, Wygnanski1973,
  Sreenivasan1986, Darbyshire1995, Durst2006, Nishi2008, Duguet2010b}.

\begin{figure}
\centering	
 \includegraphics[width=0.995\linewidth]{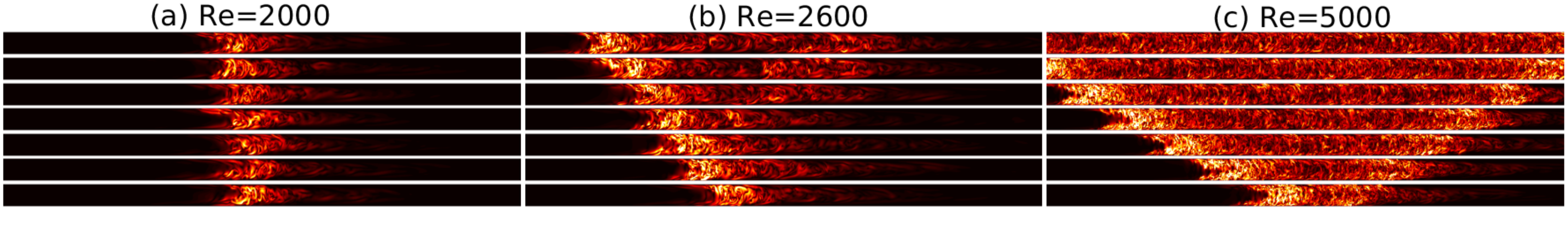}
 \includegraphics[width=0.95\linewidth]{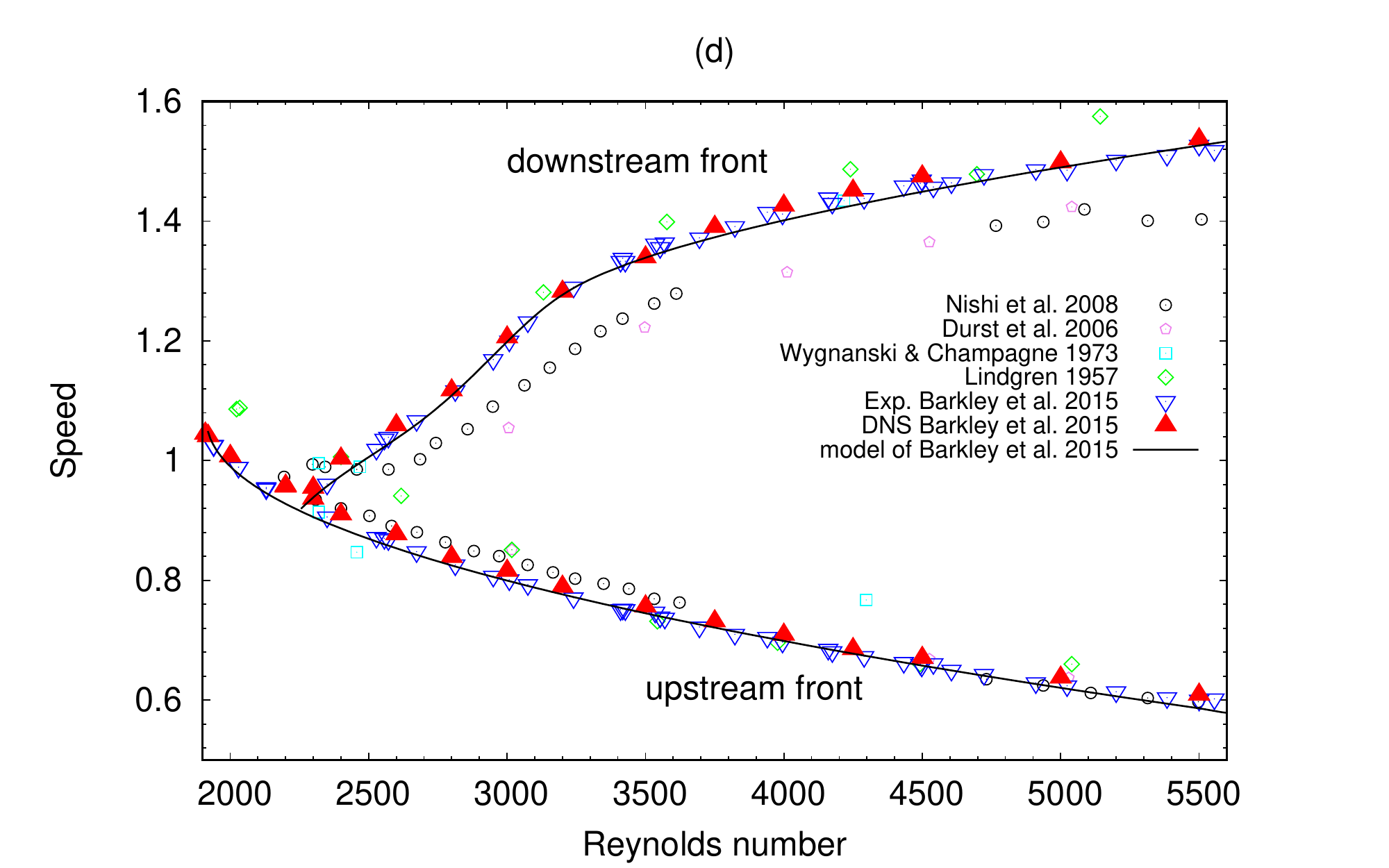}
 \caption[Shape of puffs and slugs]{\label{fig:puff_slug} Temporal evolution of
  (a) a puff at $Re=2000$, (b) a slug at $\Rey=2600$, and (c) a slug at
  $Re=5000$.  The flow is from left to right, and turbulence is visualised by
  the transverse turbulent fluctuations $q$ \DBfinal{(defined in
    \S\ref{sec:prelims})} in a frame of reference co-moving  \BSrevision{at the 
   average of the upstream and downstream front speeds}. 
  The length scale in the vertical (radial) direction is stretched
  by a factor of 2 for better visualisation. 
  Dark areas correspond to small fluctuations and bright areas correspond to large fluctuations. 
  Time evolves in the upward direction and panels are separated by $10D/U$ in (a, c) and by
  $100D/U$ in (b), where $U$ is the bulk velocity and $D$ the pipe
  diameter. (d) Front speeds as a function of Reynolds number taken from the
  literature as indicated.}
\end{figure}

In more detail, puffs are arrowhead-shaped structures with a sharp upstream
front and a diffusive downstream front (see
figure~\ref{fig:puff_slug}a). Within a puff, individual fluid parcels do not
persist in a turbulent state.  Rather turbulence is generated at the upstream
front and then decays continuously as fluid parcels pass to the downstream
front \citep{Rotta1956,Wygnanski1975,Darbyshire1995,vanDoorne2009,Hof2010}.
In contrast, at high Reynolds numbers slugs have a spatially extended bulk
region between the upstream and downstream fronts, and in the bulk region
turbulence shows no significant spatial variation, indicating that the
interior part of slugs is in a persistent turbulent state (see
figure~\ref{fig:puff_slug}c).  In their seminal experimental work
\citet{Wygnanski1973} investigated the energy budget of slugs. Their
measurements above Reynolds number 4200 (at $\Rey = 4.2\times 10^3$, $\Rey =
2\times 10^4$, and $\Rey = 2.32\times 10^4$ based on the bulk velocity $U$ and
pipe diameter $D$) showed that the upstream and downstream fronts of slugs
have a similar, well-defined structure. They observed that within the bulk
region there is sufficient turbulent kinetic energy production to sustain
turbulence and hence that the bulk region corresponds to that of fully
turbulent pipe flow.

\cite{Duguet2010b} conducted detailed direct numerical simulations of slug
formation and noted that slugs take multiple forms as manifested by different
downstream front structures.  At moderate Reynolds numbers slugs have
diffusive downstream fronts, not 
unlike the downstream fronts observed for
puffs (see figure \ref{fig:puff_slug}b), whereas at high Reynolds numbers
the downstream fronts are sharper, with a well-defined structure similar
in intensity to the upstream fronts (see figure \ref{fig:puff_slug}c).
This variation in the structure of downstream fronts can be clearly seen in
earlier experimental work \cite[e.g.][]{Nishi2008}, but \cite{Duguet2010b} are
the first to have noted the significance of the different front types.
\cite{Barkley2015} referred to the diffusive form of the downstream fronts as
weak fronts and the sharper form as strong fronts. The corresponding slugs are
thereby called weak and strong slugs, respectively.

\BSrevision{The origin of} the various structures 
has recently been elucidated with the guide of an advection-reaction-diffusion model
\citep{Barkley2011a,Barkley2015}. Figure~\ref{fig:puff_slug}(c) illustrates
this sequence by showing the speeds of upstream and downstream fronts as a
function of Reynolds number. Points show measurements from a variety of past
studies. The up and down triangles are from recent experiments and direct
numerical simulations aimed at accurately resolving the details of front
speeds within the transitional regime. The corresponding model analysis is
shown with solid curves.
At low Reynolds numbers ($\Rey\lesssim2250$) turbulence is localised in the
form of puffs and hence the upstream and downstream front speeds are
identical. Both front speeds decrease with increasing $\Rey$.  As $\Rey$ is
increased above $\Rey\simeq 2250$, the downstream front speed begins to
abruptly increase with $\Rey$, thereby deviating from the upstream front
speed. This marks the onset of expanding turbulent slugs.  Within the model
analysis, this transition from puffs to slugs is understood as a change from
excitability to bistability.  Initially, for $\Rey$ only slightly above $\Rey
\simeq 2250$ turbulent slugs take the weak form and the expansion rate of the
turbulent structure is modest.  \DBnew{ (See
  figure~\ref{fig:puff_slug}(b) and note that the time scale here is 10
  times larger than that of the other two visualisations.)}
At high Reynolds numbers 
\DBfinal{expansion is much more rapid and }
the slugs take the strong form with \BSrevision{sharp fronts at
both upstream and downstream ends} (figure~\ref{fig:puff_slug}c).  For a
comprehensive discussion on the model and theoretical perspective of the route
to turbulence in pipe flow, see \citet{Barkley2016}.

The focus of the current work is the distinction between weak and strong
slugs.  Although the dynamics of weak and strong fronts was described within a
model system, further study of real pipe flow is essential to elucidate the
physical mechanisms distinguishing them. The issues are the
following. Firstly, the model is only one dimensional, whereas in reality
fronts are evolving in a highly complicated fashion and are spatially
convoluted \citep{Holzner2013}.  This complexity results from three
dimensionality of the interface and intrinsic turbulent fluctuations and that
cannot be fully captured by the one-dimensional model. Hence it is important
to compare model fronts with those from full simulations of the Navier--Stokes
equations.
Secondly, and more importantly, the model assumes certain physical properties
of turbulent pipe flow that we establish as facts for the first time in the
present work.  Specifically, here we carry out extensive direct numerical
simulations (DNS) of the Navier--Stokes equations to analyse the physics of
the laminar-turbulent fronts of slugs.  Through the analysis we elucidate the
key distinction between weak and strong fronts, both in terms of the kinetic
energy budget across the fronts and in terms of front motion relative to the
advection speed turbulent structures within the bulk.

\section{Characterisation of front types}
\label{sec:front_shape} 

\subsection{Preliminaries}
\label{sec:prelims} 

\DBfinal{
For the numerical simulations of the Navier-Stokes equations, we have used the pipe flow code of
\citet{openpipeflow}, based on the description in \cite{Willis2009}.  Simulations are
started from a single localised disturbance in the form of a turbulent puff.
The flow is then left to evolve naturally into an expanding slug. In order to
obtain statistics, we generate a sample of slugs (typically 20 at each $Re$)
by using different initial disturbances. \BSrevision{The numerical resolutions for the simulations performed here
are the same as in Table~1 of the Extended Data of \citet{Barkley2015}.}

In principle the simulations could be initiated in multiple ways, for example,
by a forcing that mimics wall injection in the laboratory experiments 
(e.g. \citet{Darbyshire1995,Hof2003,Han2000,Reuter2004,Mellibovsky2007}), or by perturbing laminar flow with localised
pair of streamwise rolls (e.g. \citet{Mellibovsky2009}).
%
%
%
In our case, the initial conditions are all taken to be snapshots from
turbulent puff simulations at a fixed Reynolds number ($\Rey=2000$), so that
initial conditions resemble the snapshots shown in
figure~\ref{fig:puff_slug}(a). In this way all initial conditions are already
fully nonlinear turbulent states of nearly identical streamwise extent and
amplitude.
Such initial conditions always transform rapidly into slugs in simulations at
higher Reynolds number.  The initial transients corresponding to the
transformation into slugs are excluded from the analysis. Then 150 to 500
snapshots of each expanding slug are recorded. (We used a $180D$-long pipe
with periodic boundary conditions in the streamwise direction for all
simulations. The number of snapshots depends on how quickly the slug length
grows and fills the whole domain. Hence the number decreases with increasing
$\Rey$.)  For further details, see \citet{Barkley2015}.
For the phase-plane plots that appear in \S\ref{sec:sleeping_cap}, we
re-analyse the same data set generated by \cite{Barkley2015} who used the
data to obtain the front speeds shown in figure~\ref{fig:puff_slug}(d).
For the the energy budget analysis in \S\ref{sec:energy} and for the advection
speeds computations in \S\ref{sec:advection}, further computations have been
performed using the same computational techniques. 

} 

From the DNS data, we use
$$
q(z):=\sqrt{\int\int(u_r^2+u_\theta^2)rdrd\theta}
$$ 
as an indicator of the local turbulence intensity along the pipe axis, where
$r$, $\theta$, $z$ are radial, azimuthal, and axial coordinates, and $u_r$ and
$u_\theta$ are the radial and azimuthal velocity components,
(non-dimensionalised by the bulk velocity $U$). 

As proposed by \citet{Barkley2011a} the turbulence intensity, $q$, and the
axial centreline velocity, $u=u_z(r=0)$, provide a convenient and compact
characterisation of the state of the flow. For fully developed laminar flow
$q=0$ and $u=2$, while for a turbulent flow, $q>0$ and $u<2$. The lower value
of $u$ in the turbulent case results from a plug-like velocity profile
corresponding to a lower mean shear in the turbulent state.
%
%

\begin{figure}
\centering
 \includegraphics[width=0.95\linewidth]{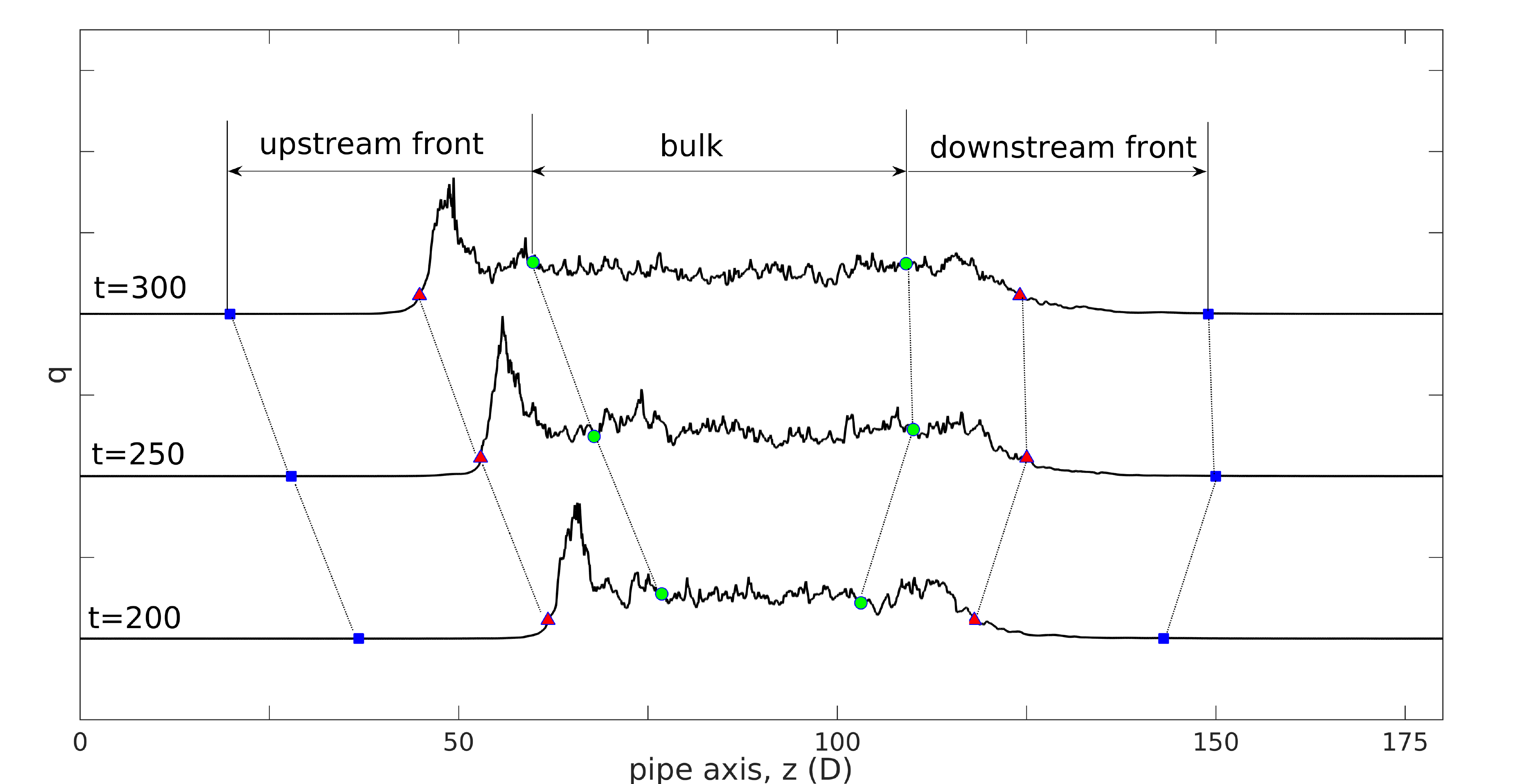}
\caption[sleepingcap plot calculation]{\label{fig:sleepingcap_calculation}
  Division of a slug into upstream front, bulk and downstream front.  A slug
  at $\Rey$=2800 is plotted at three time instants in a reference frame
  co-moving with the bulk flow. The flow is from left to right.
  \BSrevision{Red triangles mark the position of the upstream and downstream fronts given by 
  the cutoff $q=0.02$.  The bulk region is bounded between green circles, which are located
  15$D$ to the inside relative to the front positions (red triangles). Blue squares 
  mark the beginning of the fronts and are located 25$D$ to the outside relative to the front 
  positions. With these choices the fronts span 40$D$ in length.} 
  }
\end{figure}

Because turbulent slugs \DBfinal{are not fixed structures but}
expand over time, for our analysis it is necessary to
divide a turbulent slug into three parts: upstream front, {\it bulk}, and
downstream front (see figure~\ref{fig:sleepingcap_calculation}).  
A position for the laminar-turbulent fronts can be defined by setting a
threshold in $q$ above which the flow is considered as turbulent. 
We have used the threshold value $q=2\times10^{-2}$,
but we have verified that none the results that are to follow are sensitive to
the precise value of the threshold, for a substantial range of values.
The upstream and downstream front regions are then taken to be of fixed
streamwise extent ($40D$), while the size of the bulk region varies with time.
The relationship between $q$ and $u$ is numerically obtained separately in
each of the three regions. To remove the effect of fluctuations and determine
the mean dynamics of the fronts, we compute average over time (the
  saved snapshots) and the ensemble of runs.

\subsection{Fronts in the $u$-$q$ phase plane}
\label{sec:sleeping_cap}

\begin{figure}
\centering
 \includegraphics[width=0.95\linewidth]{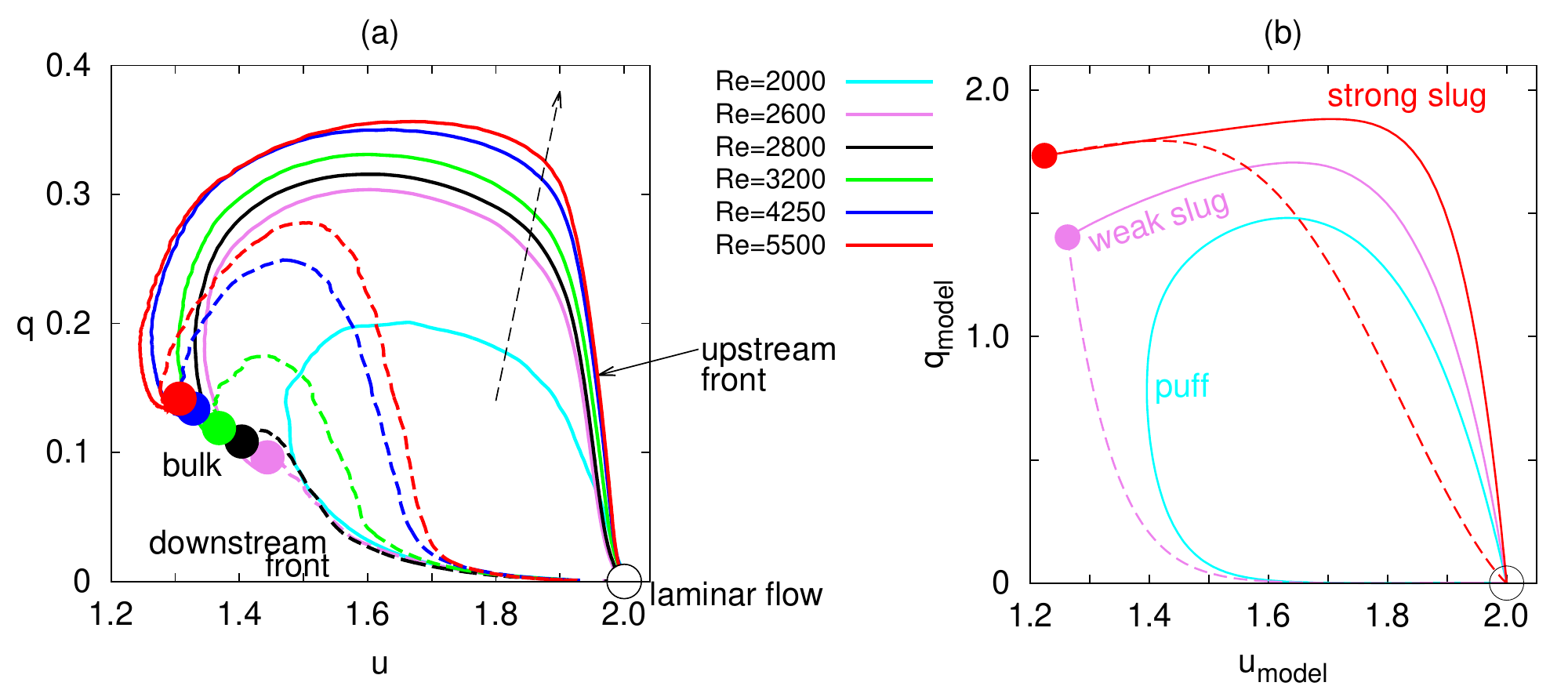}
\caption[fronts in phase space]{\label{fig:sleeping_cap} 
(a) The structure of turbulent-laminar fronts from DNS seen as trajectories in
  the $(u,q)$-phase plane.  Upstream fronts for slugs are shown with solid
  lines while downstream fronts are shown as dashed lines. Filled circles
  correspond to the bulk region of the slug. \BSrevision{The dashed arrow 
  indicates the direction of increasing Reynolds number.} The turbulent puff at
  $\Rey=2000$ does not have a bulk region and for this case we do not
  distinguish upstream and downstream fronts with different line types.
  Trajectories form counterclockwise loops in the phase plane as one travels
  in space from the upstream to the downstream along the pipe.  
  Phase-space trajectories have been obtained by averaging over time and over
  multiple runs at each $\Rey$ (typically 20 runs with 150 to 500 snapshots
  per run, depending on $Re$).
  (b) For comparison, the structure of
  turbulent-laminar fronts from a model of turbulent pipe flow
  \citep{Barkley2016}. The trajectories are taken
  directly from \cite{Barkley2016} figure~20.
}
\end{figure}

Figure~\ref{fig:sleeping_cap}(a) shows phase-space plots for turbulent slugs
at several values of $\Rey$.  A trajectory in the $(u,q)$-plane corresponds to
a spatial traverse through the structure. Counterclockwise closed loops in the
phase plane result from starting from the upstream parabolic laminar flow,
passing through the turbulent slug, and then returning back to downstream
parabolic flow.

Referring to figure~\ref{fig:sleeping_cap}(a), starting from the parabolic
laminar flow on the upstream side, there is a sharp increase in turbulence
intensity $q$ across the upstream front while the centreline velocity $u$
remains almost unchanged.  This is as expected since the modification of the
mean shear must respond to the turbulent fluctuations. This rise in $q$ occurs
both for slugs and for puffs.  
\DBfinal{For puffs the rise is rather moderate (see the cyan curve for $\Rey=2000$ in
  figure~\ref{fig:sleeping_cap}a), while for slugs, the rise is both larger
  and steeper. For slugs, the value of $u$ decreases by less than 5\% during the
  rise of $q$, in agreement with the measurements of
\cite{Wygnanski1973}.}
%
%
Following this sharp jump in $q$, the centreline velocity $u$ decreases,
corresponding to a blunting of the shear profile in response to the high level
of turbulence excitations. Subsequently, both $u$ and $q$ gradually level off
in the bulk part of the structure shown with points in the phase plot.
The constant levels of $q$ and $u$ in the bulk indicate that turbulence
production and dissipation are in balance. In fact, the values of $u$ and $q$
indicated by points are identical to those obtained from simulations of
\DBfinal{fully turbulent pipe flow at the corresponding Reynolds
  number.}

While the \DBnew{structure} of the upstream front does not vary much with
$\Rey$, the \DBnew{structure} of the downstream front does.
The distinction between weak and strong downstream fronts is quite evident 
in the phase-plane plot. 
At $\Rey=2600$ and $\Rey=2800$, as one traverses the downstream front, $q$
decreases almost monotonically from the bulk region towards zero at the
downstream side of a slug. These are the weak downstream fronts.
After the drop in $q$, $u$ slowly increases in the absence of turbulence
and approaches 2.0 (the value in the parabolic laminar flow).
%
%
In contrast, for slugs at $\Rey\gtrsim 3200$, the shape of the downstream
front in the phase plane is similar to that of an upstream front. These are
strong downstream fronts.
Here $q$ does not drop directly from the bulk, rather it first shoots upward
to a much higher level while $u$ slightly increases, followed by a sharp drop
to a very low level while $u$ hardly changes. Then $q$ vanishes while $u$
slowly approaches to 2 in the laminar flow on downstream of the slug.

As $\Rey$ increases, the overshoot of $q$ across the downstream front becomes
increasingly sharp and the shape of the curve $q(u)$ becomes very close to
that of the upstream front. 
(The upstream front is always of strong type.)
This is in agreement with \citet{Wygnanski1973},
who measured the streamwise velocity profile of upstream and downstream fronts
for $Re>4200$ in an $r$-$z$ pipe cross-section and showed that they were
similar. 

\DBmaybe{ For comparison, in figure~\ref{fig:sleeping_cap}(b) we show phase
  plane trajectories for a model puff, weak slug, and strong slug typical of
  those from the model of pipe flow proposed in \cite{Barkley2015} and
  discussed in detail in \cite{Barkley2016}. While the model captures the
  qualitative essence of the three states (a closed loop for a puff, a
  monotonic decay of $q$ for a weak downstream front and increase of $q$
  before dropping for a strong downstream front), it is clear that the model
  fails to capture quantitatively the structure of fronts from full DNS. The
  fact that the model value of $q$ do not agree with those from DNS is not
  significant, as this could be accounted for with some straightforward
  rescaling. What is significant is that the strong overshoots in $q$ observed
  in the DNS are missed by the model. This shows that further work is needed
  to obtain a model that quantitatively captures the fronts in turbulent pipe
  flow.  }

\subsection{Dynamic switching between weak and strong fronts}
\label{chp:change_front}

\begin{figure}
\centering
 \includegraphics[width=0.98\linewidth]{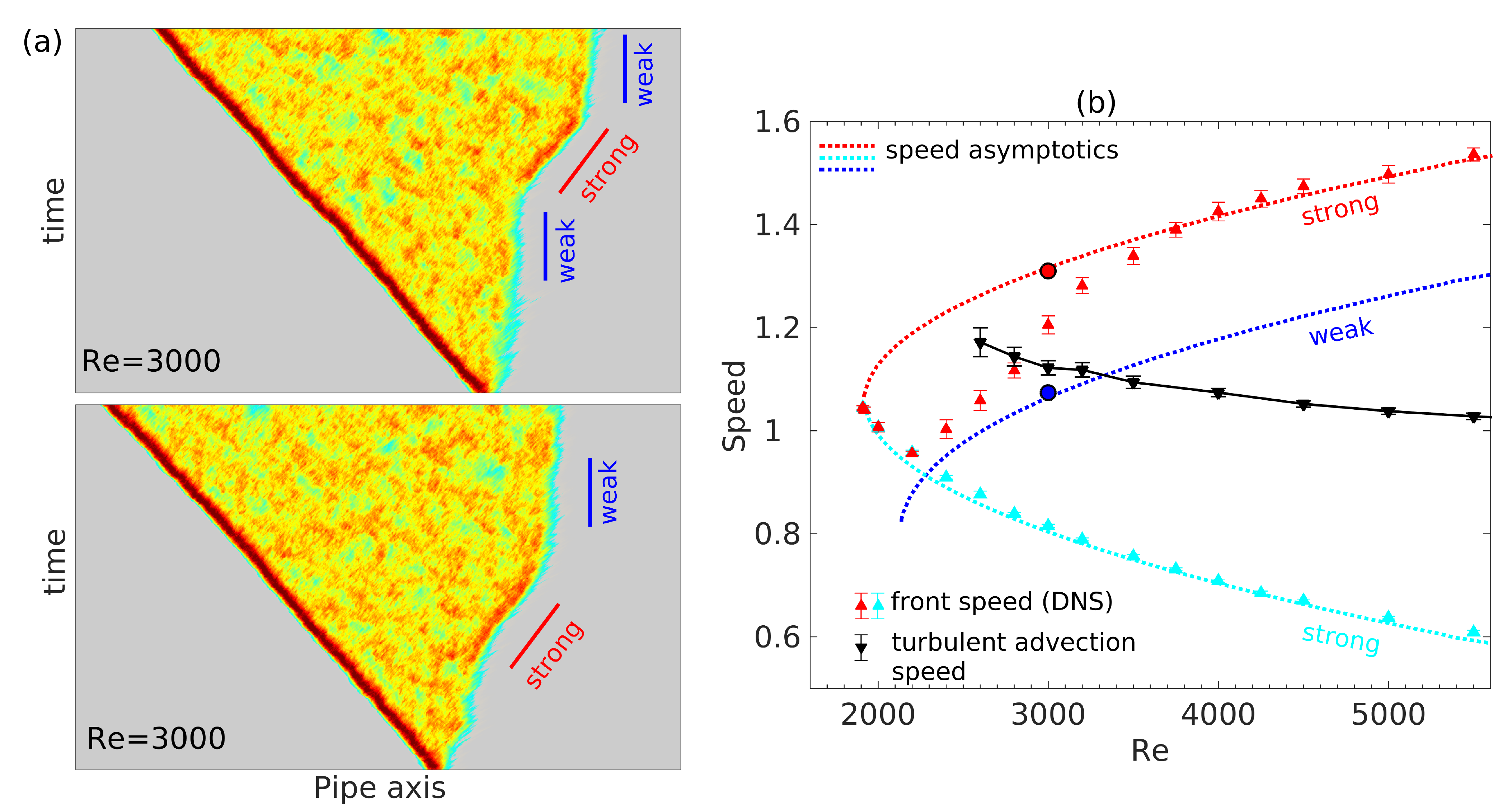}
 \caption[Switch between weak and strong fronts]{\label{fig:Re3000}
   Switching between weak and strong front at
  $\Rey=3000$. (a) Space-time diagrams showing the colourmap of $q$ for two 
  typical slugs in a moving frame of reference with the speed of 1.06$U$ 
  (the speed of the weak front at this Reynolds number). 
  \BSrevision{Grey areas correspond to small fluctuations and dark red areas correspond to large fluctuations.}
  The bars show the slopes, i.e.~the \BSrevision{inverse of} the speeds, 
  of transient weak and strong fronts \BSrevision{in this frame of reference}.
%
\DBfinal{
   (b) Front speed as a function of Reynolds number from DNS (up-triangles) and
   from an asymptotic analysis of fronts in a model system (dotted lines),
   reproduced from \citet{Barkley2015}.
}
   The circles show the speeds of the transient weak and
   strong fronts highlighted in (a). Black down-triangles show the mean turbulent
   advection speed in the bulk region of slugs computed using
   equation~\eqref{equ:adv_speed} \DBfinal{and discussed in 
   \S\ref{sec:advection} }
}
\end{figure}

The preceding description of the fronts is based on long-time and ensemble
averaging. However, the intrinsic fluctuations that are smoothed out by 
averaging are of considerable interest as they 
can result in strong deviations from the mean
behaviour. Our simulations show that in fact substantial changes in the
characteristics of the downstream front of slugs can occur during the
evolution of a single turbulence structure. 
The space-time plots of two turbulent slugs at $\Rey=3000$ in
figure~\ref{fig:Re3000}(a) illustrate the type of switching between weak and
strong downstream fronts that is frequently observed.
\DBnew{Such switching behaviour has not been reported previously and was not
observed in modelling studies.}
The speeds of these two front types have been determined separately and are
plotted on the speed diagram of figure~\ref{fig:Re3000}(b) as separate
circles. They lie almost perfectly on the weak and strong speed asymptotics
predicted by the theory of \citet{Barkley2015}, \BSrevision{who considered the limit of 
infinitesimally thin fronts. In this limit $q$ is a discontinuous function of $u$ 
instead of a steep but continuous function as shown in figure~\ref{fig:sleeping_cap}(a)}. 
As the front transiently takes one of the two forms, the ensemble average speed 
(the red up-triangle at $\Rey=3000$ in figure\ref{fig:Re3000}(b)) sits between the 
two asymptotic values. 
Our DNS data show that for $2800 < \Rey < 3200$ dynamic switches in downstream
front type are often observed. 
Outside of this range we find almost entirely
constant downstream front types, weak fronts at low $\Rey$ and strong fronts
at high $\Rey$.
%
%
%
The marked differences in propagation speed and turbulent intensity between the two 
front types calls for a deeper investigation of the physical distinction between them, 
which we address in the following sections.

\section{Energy budget at the fronts}
\label{sec:energy}

\citet{Wygnanski1973,Wygnanski1975} analysed the turbulent
kinetic energy of their experimental data, focusing on the radial
distribution and the balance of all components of the energy budget.
Hot-wire measurements were made of the three velocity
components in a $(r,z)$-plane, with limited spatial resolution (around 
$\Delta r$=0.025$D$, i.e., 20 points on the radius), and hence they had to rely on 
approximations to obtain all the terms in the budget. From the DNS performed 
in this work we have access to temporally and spatially resolved data, which 
motivates us to revisit their turbulent kinetic energy budget in order to shed 
light on the origin of the different turbulent intensity profiles at weak and 
strong fronts.

Here we use the Reynolds decomposition of the velocity field $\boldsymbol
u=\bar {\boldsymbol u} + \boldsymbol u'$ where $\boldsymbol u$ is the
total velocity, $\bar{\boldsymbol u}$ is \BSrevision{the mean velocity field 
averaged} over time and over the homogeneous
(azimuthal) direction, and $\boldsymbol u'$ is the fluctuation with respect to
the mean. The equation of the mean turbulent kinetic energy $k=\frac{1}{2}\overline{
\boldsymbol u'\cdot \boldsymbol u'}$ reads 
\citep[][pp.123-128]{Pope2000}:
\begin{equation}\label{equ:energy_budget}
\frac{\bar{\text{D}}k}{\bar{\text{D}}t}+\nabla\cdot\boldsymbol T=P-\epsilon,
\end{equation}
where $\frac{\bar{\text{D}}}{\bar{\text{D}}t}=\frac{\partial}{\partial
  t}+\bar{\boldsymbol u}\cdot\nabla$ is the material derivative
associated with $\bar{\boldsymbol u}$, and
\begin{equation}\label{equ:flux_term}
T_i=\frac{1}{2}\overline{ u_i'u_j'u_j'} + \overline{ u_i'p'} - \frac{2}{Re}\overline{ u_j's_{ij}}
\end{equation}
is the mean energy flux due to diffusion and the work by fluctuating pressure $p'$
through the surface of a fixed control volume. Here the Einstein summation
convention is used and
\begin{equation}
s_{ij}=\frac{1}{2}\left( \frac{\partial u_i'}{\partial x_j}+\frac{\partial u_j'}{\partial x_i}\right)
\end{equation}
is the fluctuating rate of strain. The production term and dissipation term are 
\begin{equation}\label{equ:P_E_terms}
P=-\overline{ u_i' u_j'} \frac{\partial \overline {u_i}}{\partial x_j},~~~~~~ \epsilon=\frac{2}{Re}\overline{ s_{ij}s_{ij}}.
\end{equation}

We note that \citet{Wygnanski1973}  defined the dissipation as
\begin{equation}
\tilde\epsilon=\frac{1}{Re}\overline{\frac{\partial u_i'}{\partial x_j}\frac{\partial u_i'}{\partial x_j}}- \frac{1}{2}\overline{\frac{\partial^2(u_i'u_i')}{\partial x_j\partial x_j}}
\end{equation}
which does not fully account for the viscous dissipation in turbulence and
mixes viscous dissipation and energy flux (transport) due to diffusion. As a
result their analysis underestimated dissipation.

\begin{figure}
\centering
 \includegraphics[width=0.95\linewidth]{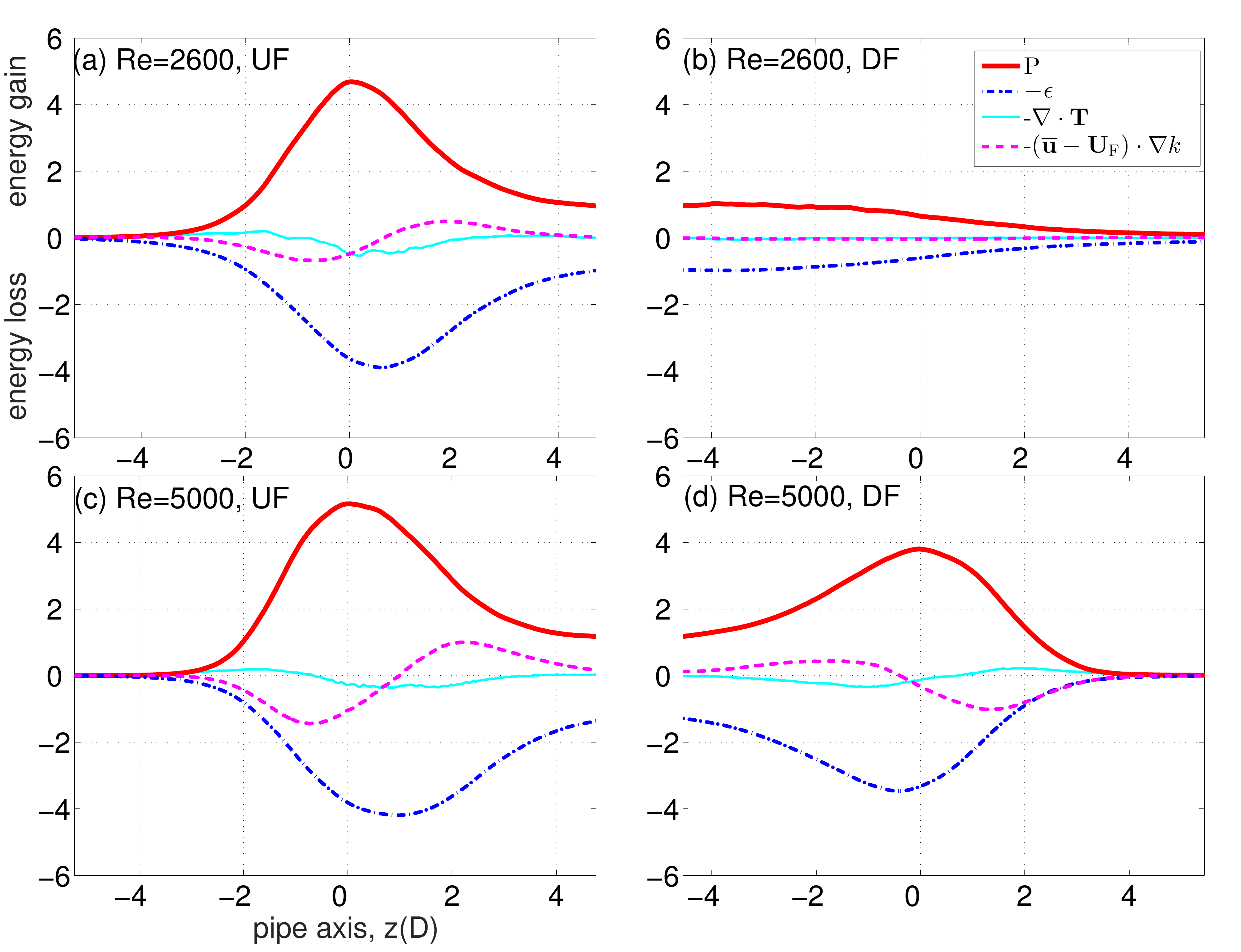}
\caption[Energy budget]{\label{fig:energy_budget} Kinetic energy budget
  (integrated over pipe cross-section) in the frame of reference co-moving
  with the fronts. Production (red bold line), dissipation (blue dash-dotted line), 
  energy flux (cyan thin line), and convection (violet dashed line)
  are normalised by the mean turbulence production rate in the bulk. (a, b)
  Upstream front (UF) and downstream front (DF) of a slug at $\Rey=2600$.  (c,
  d) Upstream front and downstream front of a slug at $\Rey=5000$. In (a, c,
  d) the peak of the production term is positioned at $z=0$. At each Reynolds
  number, an ensemble of about 500 snapshots is used to perform the energy
  budget analysis.}
 \end{figure}

In the frame of reference co-moving with the fronts at velocity 
$\boldsymbol U_{\text{F}}$, the equation for the mean
kinetic energy of the turbulent fluctuations reads
\begin{equation}\label{equ:energy_balance_advection_frame}
\frac{\partial k}{\partial t}=P-\epsilon -(\bar{\boldsymbol u} - \boldsymbol U_{\text{F}})\cdot\nabla k - \nabla\cdot\boldsymbol T=0,
\end{equation}
as the fronts are in statistical equilibrium in this frame of reference.

Figure \ref{fig:energy_budget} shows the cross-sectionally integrated energy
budget for turbulent fronts at $\Rey=2600$ (a, b) and $\Rey=5000$ (c, d). In
the figure, laminar flow is on the left for the upstream front (a, c), and on
the right for the downstream front (b, d). In the formulation of
\citet{Pope2000} used here, production and dissipation are the source and sink
terms in the energy budget and our calculation shows that they are comparable
in magnitude at all axial positions. This is in contrast to the calculation of
\citet{Wygnanski1973}, who reported that dissipation is orders of magnitude
smaller than production at the fronts (see Table~1 and 2 in their paper). We
believe that this discrepancy is due to their definition of the
dissipation term.

At strong fronts (a, c, d), as one looks into the fronts from the laminar
side, turbulence production first increases sharply, whereas the increase in
dissipation is significantly delayed with respect to the production. This is
due to the fact that it takes time for the dissipation, which acts at small
length scales, to take effect after the formation of large eddies at the front
\citep{Wygnanski1973}. This is a clear indication that there are large eddies
at the strong fronts extracting energy from the adjacent laminar flow (see
figure~\ref{fig:snaps} and online supplementary movies). As one moves
further towards the bulk, the energy production rate decreases and dissipation
outweighs production. As a result the turbulence intensity decreases and the
strong front eventually manifests an intensity peak (see
figure~\ref{fig:sleeping_cap} and figure~\ref{fig:Re3000}). Production and
dissipation eventually level off in the bulk and come into balance, while the
energy flux and convection vanish as the equilibrium (fully developed)
profile is reached. 
All terms in the energy budget for upstream and strong downstream fronts
exhibit essentially the same streamwise variation, (apart from the obvious
streamwise reflection symmetry).  This in turn is responsible for their very
similar shape in the $u$-$q$ phase plane shown in
figure~\ref{fig:sleeping_cap}.
%

At weak fronts no significant delay of dissipation with respect to production
can be observed. Here energy production and dissipation balance each other
over any cross-section (see figure~\ref{fig:energy_budget}(b)), and there is
no significant transport of energy along the streamwise direction. This
implies that in weak front turbulence is locally in equilibrium, which results
in the absence of peak in the turbulence intensity (see
figure~\ref{fig:sleeping_cap} and figure~\ref{fig:Re3000}).

\section{Fronts in the frame moving at the bulk turbulent advection speed}\label{sec:advection}

\begin{figure}
\centering
 \includegraphics[width=0.99\linewidth]{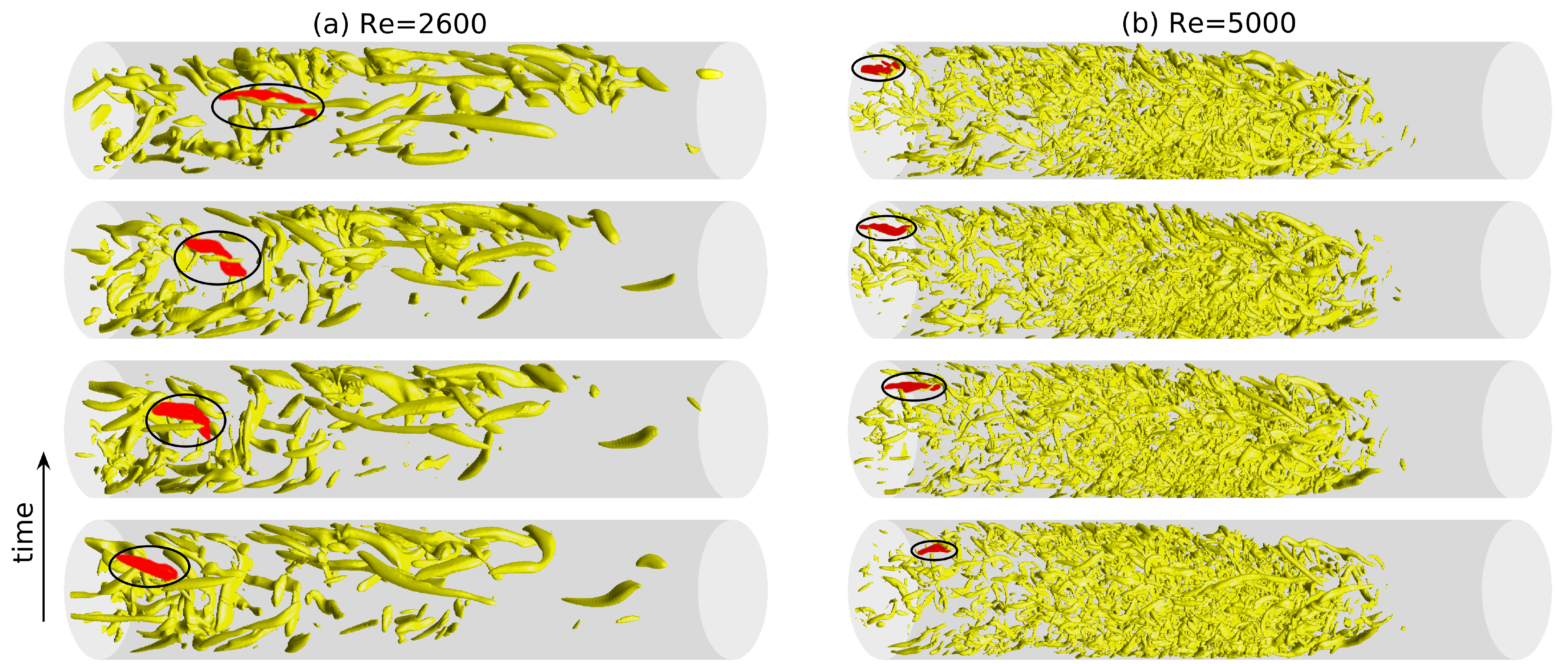}
\caption[Advection of vortices at weak and strong fronts]{\label{fig:snaps}
  \DBfinal{Advection of vortices near weak and strong fronts.}  (a) Weak front
  at $\Rey=2600$, (b) strong front at $\Rey=5000$.  Vortices, visualised by
  using the $\Lambda_2$ criterion, are shown in the frame of reference
  co-moving with the downstream front (time evolves in the upward direction).
  The bulk turbulent region is on the left and the downstream laminar flow is
  on the right.  In each case one vortex is highlighted by red colour and a
  circle and is tracked in time. In (a) the time separation between two
  consecutive panels is $1.5D/U$, whereas in (b) it is $0.625D/U$.}
\end{figure}

The energy budget analysis shows that weak and strong fronts can be
clearly distinguished by the streamwise variation in the 
turbulence production and dissipation. 
%
%
To shed light on the origin of this difference, we here investigate the
dynamics of turbulent vortices near downstream fronts.
Figure~\ref{fig:snaps}(a)--(b) show isosurfaces of $\Lambda_2$ at four time
instants in a frame co-moving with the downstream front at $Re=2600$ and
$5000$, respectively.
In each case, one of the vortices has been coloured red and circled so as to
highlight its motion relative to the front. In the $Re=2600$ case, the vortex
is generated within the turbulent core. Because it travels faster
\DBfinal{downstream} than the
downstream front, it moves towards the front. Eventually it will abandon the
turbulent slug at the front. 
Similar relative motion of the vortices with respective to the fronts for puffs was also
observed in \citet{Hof2010}.
By contrast, at $Re=5000$ the vortex
is generated at the downstream front and travels away from the front and into the
turbulent bulk. The dynamics of these two specific vortices is representative
of advecting structures at their respective Reynolds numbers (see
supplementary online movie), and suggests that, in addition to the energy
budget, weak and strong front can be distinguished by the speed at which
vortices are advected relative to the downstream front speed.

In wall-bounded shear flows the advection speed of turbulent structures
(vortices and streaks) depends on their size and distance from the wall
\citep{Alamo2009,Pei2012}. However, \citet{Alamo2009} showed that only small
vortices exhibit a clear radial dependence in their advection speed, whereas
relatively large vortices are in fact advected at a rather uniform speed (also
evidenced in \citet{Duguet2010b}). Recently, \citet{Alamo2009,Kreilos2014} proposed a
method to define and remove the mean advection speed of turbulence in parallel
shear flows. 

Starting from the Navier--Stokes equation
\begin{equation}\label{equ:navier-stokes}
\partial_t\boldsymbol{u}(r,\theta,z,t) = -\boldsymbol u \cdot\nabla
\boldsymbol u -\nabla p + \frac{1}{Re}\nabla^2 \boldsymbol u 
\end{equation}
where $\boldsymbol u$ is the velocity field and $p$ the pressure.
Letting $\boldsymbol{f}(\boldsymbol{u}(r,\theta,z,t),p,t)$ denote the
right-hand side of the Navier--Stokes equation 
equation \ref{equ:navier-stokes}, the mean turbulent
advection speed can be computed as
\begin{equation}\label{equ:adv_speed}
c =
-\frac{<\partial_z\boldsymbol{u}(r,\theta,z,t)\cdot\boldsymbol{f}(\boldsymbol{u}(r,\theta,z,t),p,t)>}
{\|\partial_z\boldsymbol{u}(r,\theta,z,t)\|^2}.
\end{equation}
Here the inner product and 2-norm of vector fields $\boldsymbol a$ and $\boldsymbol b$ are defined as
$<\boldsymbol{a}\cdot\boldsymbol{b}>=\int_V(\boldsymbol{a}\cdot\boldsymbol{b})dV$
and $\|\boldsymbol{a}\|^2=<\boldsymbol{a}\cdot\boldsymbol{a}>$. 
Note that equation \eqref{equ:adv_speed} is obtained by minimising 
$\boldsymbol{u}(r,\theta,z+\Delta z,t+\Delta t)-\boldsymbol{u}(r,\theta,z,t)$ in the discrete form
\[
<\partial_z\boldsymbol{u}(r,\theta,z,t)\cdot\left(\boldsymbol{u}(r,\theta,z+\Delta
z,t+\Delta t)-\boldsymbol{u}(r,\theta,z,t)\right)>=0,
\]
with $\Delta z=c\Delta t$. 

\BSrevision{ This method allows the calculation of an instantaneous 
mean turbulent advection speed using only a single velocity snapshot and the local time derivatives, 
therefore, has advantages compared to the methods based on calculating the velocity correlation between
flow fields with a proper time separation, which is not known a priori and introduces uncertainties
(see e.g., \citet{Kim1993,Pei2012}. It has been shown to be able to precisely obtain the advection speed of a
travelling wave solution in channel flow \citep{Kreilos2014}.}

\begin{figure}
\centering
\includegraphics[width=0.95\linewidth]{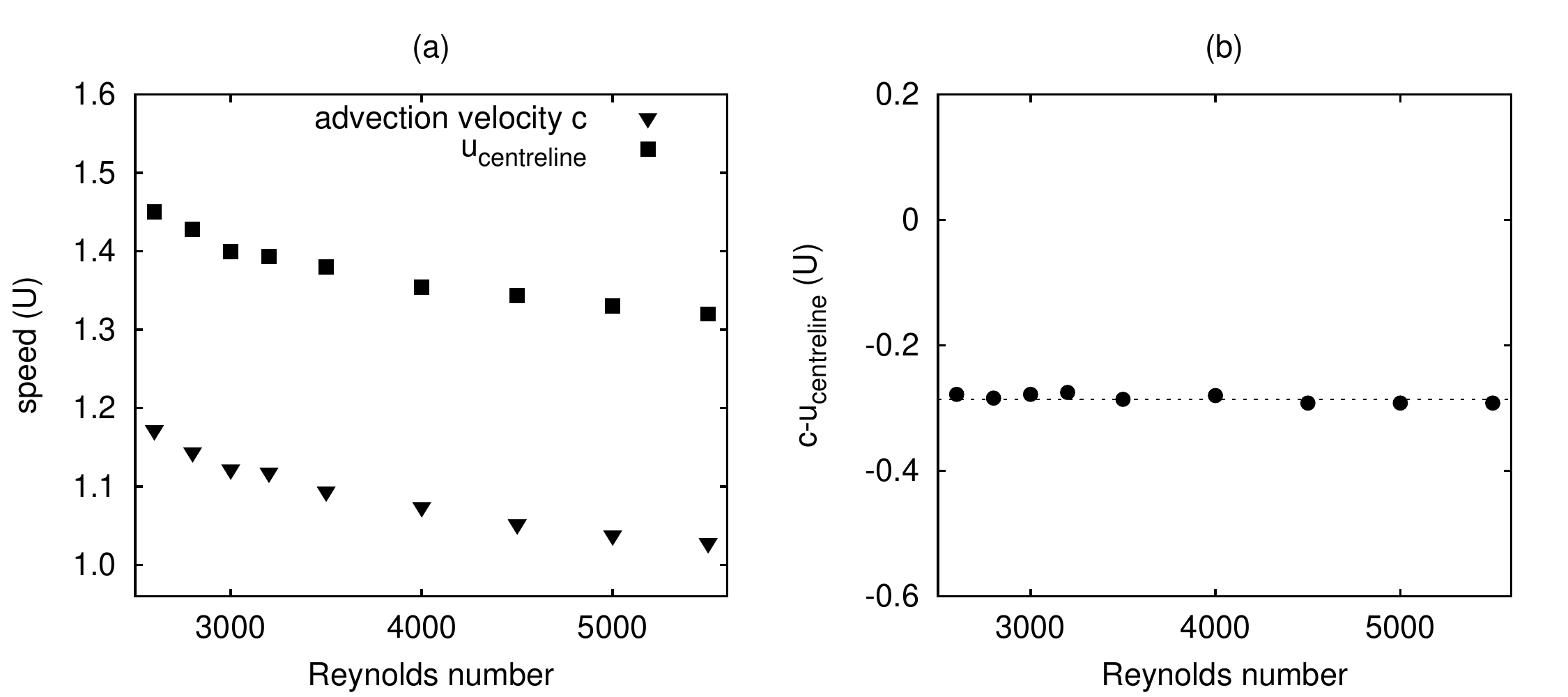}
\caption{\label{fig:adv_Vc} \BSrevision{Comparison 
between the average turbulent advection speed and 
the average centreline velocity as a function of $\Rey$. (b) Difference between the 
advection speed and the centreline velocity.}}
\end{figure}

The black down-triangles in figure~\ref{fig:Re3000}(b) \BSrevision{and figure~\ref{fig:adv_Vc}(a)} show the mean turbulent
advection speed as a function of the Reynolds number obtained with equation~\ref{equ:adv_speed}.
The computations are for \DBfinal{fully turbulent pipe flow} and give the
speed of the bulk turbulent structures.  These \BSrevision{calculations} reveal a simple
relation between the average turbulent advection speed and the average
centreline velocity in the bulk \BSrevision{(see figure~\ref{fig:adv_Vc}(b))},
\begin{equation}
c=u_{\text{centreline}}-(0.286\pm 0.008),
\end{equation}
\BSrevision{where $0.008$ is the standard deviation of the speed difference.} This relation holds in the whole range 
of Reynolds numbers investigated.
\DBnew{ It should be noted that for the purposes of modelling, 
  \cite{Barkley2015} and \cite{Barkley2016} assumed that turbulence is
  advected more slowly than the centreline velocity by a fixed constant. The
  results here demonstrate that this is in fact the case.}

It is particularly enlightening to view expanding turbulent structures (slugs)
in the frame of reference co-moving with the advection speed of the bulk
turbulence $c$. In such a frame of reference, the structures in the bulk of
slugs overall stay still, while the fronts move. 
As seen in figure~\ref{fig:Re3000}(b), at low $\Rey\lesssim 2900$, mean
turbulent advection speed $c$ is larger than the speeds of both upstream front
(cyan up-triangles) and downstream front (red up-triangles). Hence, in the frame in
which the bulk turbulent structures are stationary, both fronts move to the
left (upstream direction). New turbulent structures are generated at the
upstream front as it moves to the left. These structures then decay back to
laminar flow as the downstream front subsequently passes.
For $\Rey\gtrsim 2900$ the speed of the strong front is larger than $c$.
Hence, in the frame in which the bulk turbulent structures are stationary,
downstream fronts now moves to the right (downstream direction), 
generating new turbulent structures. 
Hence turbulent structures are now generated at both fronts and the large
eddies produced at the fronts extract energy from the adjacent laminar flow,
on both the upstream and downstream end of the slug, in agreement with the
energy budget analysis.
%
%
This strong correlation between front speed
and turbulence intensity can also be seen in the colourmap of
figure~\ref{fig:Re3000}: the strong front is red (high $q$), whereas the weak
front is yellow. From our argument it could thus be inferred that weak slugs
feature a decaying downstream front whereas strong slugs possess a
turbulence-producing downstream front, and the transition between weak and
strong fronts should occur at $\Rey\approx 2900$. Our data support this
argument and indeed below $2900$, weak fronts are (statistically) more likely
to occur, while strong fronts dominate above $2900$.

\section{Discussion and conclusion}\label{sec:discussion}

By performing and analysing extensive direct numerical simulations, we have
studied the propagation and structure of laminar-turbulent fronts in pipe flow
over a wide range of Reynolds numbers, from $\Rey = 2000$ to $\Rey =
5500$. The structure of the upstream front remains qualitatively unchanged
over the whole Reynolds number range and its dimensionless speed monotonically
decreases as $\Rey$ increases. The main feature of this front is a sharp peak
in turbulent intensity, which is due to the ability of the upstream front to
extract energy from the upstream laminar flow. As a result turbulent
production is at the upstream front much larger than in the bulk.  Dissipation
within the upstream front is similar in magnitude to production, but the peak
dissipation is reached downstream of the peak production. \DBnew{This effect
  has been missed in prior studies.} This delay arises because the energy is
transferred from the large eddies at the interface down to the smallest scales
as the turbulent structures are advected downstream.

The behaviour of the downstream front is more complex and changes drastically
as the Reynolds number increases. At low Reynolds numbers the downstream front
of slugs is weak, i.e.~consisting of a gradual relaminarisation toward the
parabolic flow profile as seen in puffs. \DBnew{Interestingly, we have found
  that } turbulence is locally in equilibrium at weak fronts: at each
streamwise location production and dissipation balance and there is no
streamwise transport of kinetic energy by diffusion or convection.  As $Re$
increases, strong downstream fronts, characterised by a peak in turbulent
intensity similar to that of upstream fronts, are found to dominate.

Weakly and strongly expanding slugs can be clearly seen in previous
experimental work \cite[e.g.][]{Nishi2008} and the distinction was first noted
by \citet{Duguet2010b}. A model analysis in the limit of sharp fronts gives a
theoretical foundation for the distinction and suggests that at large Reynolds
numbers the strong downstream front becomes the symmetric image of the
upstream front \citep{Barkley2015}. Our turbulent kinetic energy budget
analysis confirms this prediction. Our results are in line with the
conclusions of the seminal experimental studies of Wygnanski and co-workers
\citep{Wygnanski1973,Wygnanski1975}, and provide a bridge between their
analysis of the physics and the recent theoretical model of
\citet{Barkley2015}. Note however that a different definition of dissipation
was used by Wygnanski, which does not fully account for the viscous
dissipation in turbulence. As a result \citep{Wygnanski1973} underestimated
dissipation at the fronts, which led them to conclude that there is a large
net production at strong fronts.

Based on a comparison between the advection speed of turbulence in the bulk of
slugs and the speed of fronts, our study reveals the \DBnew{fundamental}
mechanism of transition between weak and strong fronts. For $\Rey\lesssim
2900$ the downstream front travels slower than the bulk turbulent advection
speed and so in the co-moving frame, turbulence relaminarises along the weak
front. In contrast, for $\Rey\gtrsim 2900$ the downstream front travels faster
than the bulk turbulent advection speed and so it can aggressively invade the
laminar flow ahead, while extracting energy from it just as the upstream front
does. In the intermediate regime around $\Rey\simeq 2900$, fluctuations of
turbulence production/dissipation, may cause temporal switching between the
two types of front (see figure~\ref{fig:Re3000}). Hence the transition from
weak to strong fronts is of statistical nature, exactly as the transition from
transient to sustained turbulence \citep{avila2011} or the onset of weakly
expanding slugs~\citep{Barkley2015}.

Our study has been restricted to pipe flow, but we believe that the front
dynamics and physical mechanisms shown here are generic to wall-bounded
turbulent flows. Experiments in square ducts display both types of fronts and
hence support this hypothesis \citep{Barkley2015}. In flows with two spatially
extended dimensions, such as channel and Couette flows, advection is possible
in the spanwise and streamwise directions \citep{Barkley2007,Duguet2013},
resulting in substantially more complex front dynamics and challenging current
modelling strategies.

\section*{Acknowledgements}

We thank Dr.~Ashley P.~Willis for sharing his spectral toroidal-poloidal pipe
flow code and Anna Guseva for reading the manuscript.  We acknowledge the
Deutsche Forschungsgemeinschaft (Project No. FOR 1182), and the European
Research Council under the European Union’s Seventh Framework Programme
(FP/2007-2013)/ERC Grant Agreement 306589 for financial
support. B.~S. acknowledges financial support from the Chinese State
Scholarship Fund under grant number 2010629145, support from the Max Planck
Society and the Institute of Science and Technology Austria (IST Austria). We
acknowledge the computing resources from Gesellschaft f\"ur wissenschaftliche
Datenverarbeitung G\"ottingen (GWDG), the J\"ulich Supercomputing Centre
(grant HGU16), the Regionalen Rechenzentrums Erlangen (RRZE) and computing
facilities at IST Austria.

\bibliographystyle{jfm}

\end{document}